\documentclass[3p,twocolumn]{elsarticle}
\pdfoutput=1
\usepackage[acronym,toc]{glossaries}
\usepackage{amsmath}
\newacronym{MIT}{MIT}{the Massachusetts Institute of Technology}
\newacronym{UW}{UW}{University of Wisconsin}
\newacronym{US}{US}{United States}
\newacronym{IAEA}{IAEA}{International Atomic Energy Agency}
\newacronym{SNF}{SNF}{spent nuclear fuel}
\newacronym{HLW}{HLW}{high level waste}
\newacronym{FEHM}{FEHM}{Finite Element Heat and Mass Transfer}
\newacronym{DOE}{DOE}{Department of Energy}
\newacronym{GENIUS}{GENIUS}{Global Evaluation of Nuclear Infrastructure Utilization Scenarios}
\newacronym{GENIUSv1}{GENIUSv1}{Global Evaluation of Nuclear Infrastructure Utilization Scenarios, Version 1}
\newacronym{GENIUSv2}{GENIUSv2}{Global Evaluation of Nuclear Infrastructure Utilization Scenarios, Version 2}
\newacronym{CNERG}{CNERG}{Computational Nuclear Engineering Research Group}
\newacronym{GDSM}{GDSM}{Generic Disposal System Model}
\newacronym{GDSE}{GDSE}{Generic Disposal System Environment}
\newacronym{GPAM}{GPAM}{Generic Performance Assessment Model}
\newacronym{FEPs}{FEPs}{Features, Events, and Processes}
\newacronym{EBS}{EBS}{Engineered Barrier System}
\newacronym{EDZ}{EDZ}{Excavation Disturbed Zone}
\newacronym{YMR}{YMR}{Yucca Mountain Repository Site}
\newacronym{EPA}{EPA}{Environmental Protection Agency}
\newacronym{PEI}{PEI}{Peak Environmental Impact}
\newacronym{VISION}{VISION}{the Verifiable Fuel Cycle Simulation Model}
\newacronym{NUWASTE}{NUWASTE}{Nuclear Waste Assessment System for Technical Evaluation}
\newacronym{NWTRB}{NWTRB}{Nuclear Waste Technical Review Board}
\newacronym{OCRWM}{OCRWM}{Office of Civilian Radioactive Waste Management}
\newacronym{UFD}{UFD}{Used Fuel Disposition}
\newacronym{DYMOND}{DYMOND}{Dynamic Model of Nuclear Development }
\newacronym{DANESS}{DANESS}{Dynamic Analysis of Nuclear Energy System Strategies}
\newacronym{CAFCA}{CAFCA}{ Code for Advanced Fuel Cycles Assessment }
\newacronym{ORION}{ORION}{ORION}
\newacronym{NFCSim}{NFCSim}{Nuclear Fuel Cycle Simulator}
\newacronym{COSI}{COSI}{Commelini-Sicard}
\newacronym{FCT}{FCT}{Fuel Cycle Technology}
\newacronym{SWF}{SWF}{Separations and Waste Forms}
\newacronym{FCO}{FCO}{Fuel Cycle Options}
\newacronym{RDD}{RD\&D}{research development and design}
\newacronym{WIPP}{WIPP}{Waste Isolation Pilot Plant}
\newacronym{ANDRA}{ANDRA}{Agence Nationale pour la gestion des D\'echets RAdioactifs, the French National Agency for Radioactive Waste Management}
\newacronym{TSM}{TSM}{Total System Model}
\newacronym{LANL}{LANL}{Los Alamos National Laboratory}
\newacronym{INL}{INL}{Idaho National Laboratory}
\newacronym{ANL}{ANL}{Argonne National Laboratory}
\newacronym{SNL}{SNL}{Sandia National Laboratory}
\newacronym{LBNL}{LBNL}{Lawrence Berkeley National Laboratory}
\newacronym{LLNL}{LLNL}{Lawrence Livermore National Laboratory}
\newacronym{NAGRA}{NAGRA}{National Cooperative for the Disposal of Radioactive Waste}
\newacronym{CUBIT}{CUBIT}{CUBIT Geometry and Mesh Generation Toolkit}
\newacronym{CSNF}{CSNF}{commercial spent nuclear fuel}
\newacronym{DSNF}{DSNF}{DOE spent nuclear fuel}
\newacronym{MTHM}{MTHM}{metric tons of heavy metal}
\newacronym{HTGR}{HTGR}{high temperature gas reactor}
\newacronym{TRISO}{TRISO}{Tristructural Isotropic}
\newacronym{MA}{MA}{minor actinide}
\newacronym{CEA}{CEA}{Commissariat \`a l'\'Energie Atomique et aux \'Energies Alternatives}
\newacronym{SKB}{SKB}{Svensk K\"{a}rnbr\"{a}nslehantering AB}
\newacronym{SINDAG}{SINDA{\textbackslash}G}{Systems Improved Numerical Differencing Analyzer $\backslash$ Gaski}
\newacronym{STC}{STC}{specific temperature change}
\newacronym{LDRD}{LDRD}{laboratory directed research and development}
\newacronym{LCOE}{LCOE}{levelized cost of electricity}
\newacronym{ABM}{ABM}{agent-based modeling}
\newacronym{COTS}{COTS}{commercial, off-the-shelf}
\newacronym{API}{API}{application programming interface}
\newacronym{RIF}{RIF}{Region-Institution-Facility}
\newacronym{GUI}{GUI}{graphical user interface}
\newacronym{HPC}{HPC}{high-performance computing}
\newacronym{HTC}{HTC}{high-throughput computing}
\newacronym{UML}{UML}{Unified Modeling Language}
\newacronym{DAG}{DAG}{directed acyclic graph}
\newacronym{MOX}{MOX}{mixed oxide}
\newacronym{ThOX}{ThOX}{thorium oxide}
\newacronym{UOX}{UOX}{uranium oxide}
\newacronym{QA}{QA}{quality assurance}
\newacronym{NQA1}{NQA-1}{Nuclear Quality Assurance - 1}
\newacronym{VV}{V\&V}{verification and validation}
\newacronym{UQ}{UQ}{uncertainty quantification}
\newacronym{ASME}{ASME}{American Society of Mechanical Engineers}
\newacronym{NEAMS}{NEAMS}{Nuclear Engineering Advanced Modeling and Simulation}
\newacronym{CI}{CI}{continuous integration}
\newacronym{DRE}{DRE}{dynamic resource exchange}
\newacronym{DESAE}{DESAE}{Dynamic Analysis of Nuclear Energy Systems Strategies}
\newacronym{MILP}{MILP}{mixed-integer linear program}
\newacronym{LP}{LP}{linear program}
\newacronym{NEUP}{NEUP}{Nuclear Energy University Programs}
\newacronym{NRC}{NRC}{Nuclear Regulatory Commission}
\newacronym{NSF}{NSF}{National Science Foundation}
\newacronym{NNSA}{NNSA}{National Nuclear Security Administration}
\newacronym{DHS}{DHS}{Department of Homeland Security}
\newacronym{SWU}{SWU}{Separative Work Unit}

\makeglossaries

\usepackage{float} 

\usepackage{lineno,hyperref}
\modulolinenumbers[5]

\usepackage{listings} 
\usepackage{color}

\definecolor{dkgreen}{rgb}{0.13, 0.55, 0.13}
\lstdefinelanguage{diff}{
  morecomment=[l][\color{blue}]{@@},     
  morecomment=[l][\color{red}]-,         
  morecomment=[l][\color{dkgreen}]+,       
  morecomment=[l][\color{magenta}]{---}, 
  morecomment=[l][\color{magenta}]{+++},
}
\definecolor{bkggray}{rgb}{0.95,0.95,0.95}
\DeclareFontFamily{\encodingdefault}{\ttdefault}{\hyphenchar\font=`\-} 

\usepackage{tabularx}
\usepackage{graphicx}
\usepackage{booktabs} 
\usepackage{microtype} 
\usepackage{xspace}
\newcommand{\Cyclus}{\textsc{Cyclus}\xspace}%
\newcommand{\Cycamore}{\textsc{Cycamore}\xspace}%
\newcommand{\Class}[1] {\texttt{#1}}%
\journal{Advances in Engineering Software}

\begin{document}
\begin{frontmatter}

\title{Fundamental concepts in the Cyclus nuclear fuel cycle simulation framework}

\author[berk]{Kathryn D. Huff\corref{corrauthor}}
\cortext[corrauthor]{Corresponding Author}
\ead{huff@berkeley.edu}

\author[wisc]{Matthew J. Gidden}
\author[wisc]{Robert W. Carlsen}
\author[austin]{Robert R. Flanagan}
\author[wisc]{Meghan B. McGarry}
\author[wisc]{Arrielle C. Opotowsky}
\author[austin]{Erich A. Schneider}
\author[usc]{Anthony M. Scopatz}
\author[wisc]{Paul P.H. Wilson}

\address[berk]{University of California - Berkeley, Department of Nuclear Engineering Berkeley, CA, 94720}
\address[wisc]{University of Wisconsin - Madison, Department of Nuclear Engineering and Engineering Physics, Madison, WI 53706}
\address[austin]{University of Texas - Austin, Department of Mechanical Engineering, Nuclear and Radiation Engineering Program, Austin, TX 78758}
\address[usc]{University of South Carolina, Nuclear Engineering Program, Columbia, SC 29201}

\begin{abstract}
As nuclear power expands, technical, economic, political, and environmental
analyses of nuclear fuel cycles by simulators increase in importance.  To date,
however, current tools are often fleet-based rather than discrete and 
restrictively licensed rather than open source. Each of these choices presents a challenge
to modeling fidelity, generality, efficiency, robustness, and scientific
transparency.  The \Cyclus nuclear fuel cycle simulator framework and its
modeling ecosystem incorporate modern insights from simulation science and
software architecture to solve these problems so that challenges in nuclear
fuel cycle analysis can be better addressed.  A summary of the \Cyclus fuel
cycle simulator framework and its modeling ecosystem are presented.
Additionally, the implementation of each is discussed in the context of
motivating challenges in nuclear fuel cycle simulation.  Finally, the current
capabilities of \Cyclus are demonstrated for both open and closed fuel
cycles.
\end{abstract}


\end{frontmatter}



\section{Introduction}


As nuclear power expands, technical, economic, political, and environmental
analyses of nuclear fuel cycles by simulators increase in importance. The
merits of advanced nuclear technologies and fuel cycles are
shaped by myriad physical, nuclear, chemical, industrial, and political
factors. Nuclear fuel cycle simulators must therefore couple complex models of
nuclear process physics, facility deployment, and material routing.

Indeed, the cardinal purpose of a dynamic nuclear fuel cycle simulator is to calculate
the time- and facility-dependent mass flow through all or part the fuel cycle.
Dynamic nuclear fuel cycle analysis more realistically supports a range of
simulation goals than static analysis \cite{piet_dynamic_2011}. Historically,
dynamic nuclear fuel cycle simulators have calculated fuel cycle mass balances
and performance metrics derived from them using software ranging from
spreadsheet-driven flow calculators to highly specialized system dynamics
modeling platforms.

To date, current tools are typically distributed under restrictive rather than open
source licenses, having been developed in industrial contexts or using 
commercial software platforms. Additionally, having
often been developed for customized applications, many possess inflexible
architectures, never having been designed to enable new features or extensions.
Finally, many model only fleet-level dynamics of facilities and
materials rather than discrete resolution of those individual agents and
objects. When the DOE-NE Fuel Cycles Technologies Systems Analysis Campaign
developed requirements necessary in a next generation fuel cycle simulator,
three main failure modes were associated with those software choices.
First, they discourage targeted
contribution and collaboration among experts. Next, they hobble efforts to
directly compare modeling methodologies. Finally, they over-specialize,
rendering most tools applicable to only a subset of desired simulation
fidelities, scales, and applications. Those three constraints were identified
as presenting significant challenges to modeling fidelity, generality, efficiency,
robustness, and scientific transparency in the field of fuel cycle analysis
\cite{huff_next_2010}.

The \Cyclus nuclear fuel cycle simulator framework and its \emph{modeling
ecosystem}, the suite of agents and other physics plug-in libraries compatible with it,
incorporate modern insights from simulation science and software architecture
to solve these problems.  These modern methods simultaneously enable more
efficient, accurate, robust, and validated analysis.  This next-generation fuel
cycle simulator is the result of design choices made to:

\begin{itemize}
\item support access to the tool by fuel cycle analysts and other users,
\item encourage developer extensions,
\item enable plug-and-play comparison of modeling methodologies,
\item and address a range of analysis types, levels of detail, and analyst sophistication.
\end{itemize}

\Cyclus is a dynamic, agent-based model, which employs a modular architecture,
an open development process, discrete agents, discrete time, and arbitrarily
detailed isotopic resolution of materials. Experience in the broader field of
systems analysis indicates that agent-based modeling enables more flexible
simulation control than system dynamics, without loss of generality
\cite{macal_agent-based_2010}. Furthermore, openness allows cross-institutional
collaboration, increases software robustness
\cite{cohen_modern_2010,fagan_design_2002}, improves
the strength and quality of results through peer review
\cite{donoho_reproducible_2009,ince_case_2012,stodden_scientific_2010,wicherts_willingness_2011,petre_code_2014}, and
cultivates an ecosystem of modeling options. This ecosystem is \emph{modular},
being comprised of dynamically loadable, interchangeable, plug-in libraries of fuel cycle component
process physics that vary in their scope, depth, and fidelity. This modularity
allows users and developers to customise \Cyclus to analyze the cases that are
of interest to them rather than any custom application the simulator was
originally developed to address. Additionally, that customizability allows
users and developers to address those cases at the level of fidelity necessary
for their application. The fundamental concepts of the \Cyclus
nuclear fuel cycle simulator capture these modern insights so that novel
challenges in nuclear fuel cycle analysis can be better addressed.

\subsection{Background}



Nuclear fuel cycle simulators drive \gls{RDD} by calculating `metrics',
quantitative measures of performance that can be compared among fuel cycle
options.  The feasibility of the technology development and deployment
strategies which comprise a fuel cycle option, the operational features of
nuclear energy systems, the dynamics of transitions between fuel cycles, and
many other measures of performance can be expressed in terms of these metrics.
For example, economic feasibility is often measured in \gls{LCOE}, a
combination of fuel and operating costs normalized by electricity generation, while
environmental performance might be measured by spent fuel volume, radiotoxicity, or
mined uranium requirements. A meta-analysis of fuel cycle systems studies 
identified over
two dozen unique quantitative metrics spanning economics and cost,
environmental sustainability and waste management impacts, safety, security and
nonproliferation, resource adequacy and utilization, among others
\cite{flicker_evaluation_2014}. With few exceptions, these metrics are derived from
mass balances and facility operation histories calculated by a fuel cycle
simulator. For example, where nuclear waste repository burden is derived from
ejected fuel masses, water pollution or land use can be derived from facility
operational histories (as in \cite{poinssot_assessment_2014}).

However, methods for calculating those metrics vary among simulators. Some
model the system of facilities, economics, and materials in static equilibrium,
while other simulators capture the dynamics of the system.  Similarly, while
some simulators discretely model batches of material and individual facilities,
others aggregate facilities into fleets and materials into streams. Some
simulators were designed to model a single aspect of the fuel cycle in great
detail while neglecting others. For example, a simulator created for policy
modeling might have excellent capability in economics while capabilities for
tracking transformations in material isotopics and the effects of isotopics on
technology performance are neglected.  The \gls{CAFCA}\cite{guerin_impact_2009}
simulator is problem-oriented in this way, having elected to neglect isotopic
resolution in favor of integral effects.

Historically, domestic national laboratories have driven development and
regulated the use of their own tools:
\gls{VISION}\cite{jacobson_verifiable_2010},
\gls{DYMOND}\cite{yacout_modeling_2005}, and
\gls{NFCSim}\cite{schneider_nfcsim:_2005,allan_guidance_2008}.  Internationally,
other laboratories have created their own as well, such as
\gls{COSI}\cite{boucher_cosi_2005,boucher_cosi:_2006,meyer_new_2009,coquelet-pascal_comparison_2011}
and ORION\cite{worrall_scenario_2007}.  Finally, some simulators initiated in a national lab setting have
been continued as propriety, industry-based simulators, such as
\gls{DANESS}\cite{van_den_durpel_daness_2009}.  Outside the national laboratories,
researchers have created new nuclear fuel cycle simulation tools when existing
tools were not available or not sufficiently general to calculate their metrics
of interest.  With limited access to the
national laboratory tools and a need to customize them for research purposes,
universities and private industry researchers have ``reinvented the wheel'' by
developing tools of their own from scratch and tailored to their own needs.
Examples include \gls{CAFCA}\cite{guerin_benchmark_2009} and
\gls{DESAE}\cite{andrianova_desae_2008,mccarthy_benchmark_2012,allan_guidance_2008}.

\Cyclus emerged from a line of tools seeking to break this practice.  Its
precursor, \gls{GENIUS} Version 1
\cite{dunzik-gougar_global_2007,jain_transitioning_2006}, originated within
\gls{INL} and sought to provide generic regional capability.  Based on lessons
learned from \gls{GENIUS} Version 1, the \gls{GENIUS} Version 2
\cite{oliver_studying_2009,huff_geniusv2_2009} simulator sought to provide more
generality and an extensible interface to facilitate collaboration.  The \Cyclus
project then improved upon the \gls{GENIUS} effort by implementing increased
modularity and encapsulation.  The result is a dynamic simulator that treats
both materials and facilities discretely, with an architecture that permits
multiple and variable levels of fidelity. Using an agent-based framework, the
simulator tracks the transformation and trade of resources between autonomous
regional and institutional entities with customizable behavior and
objectives. Each of these concepts (agent-based, resource tracking, and
regional as well as institutional entities) will be described in their own
sections (sections \ref{sec:abm}, \ref{sec:mats}, and \ref{sec:rif}
respectively).  Together, they provide a capability for extension and reuse
beyond that pursued by any existing fuel cycle simulator.

\subsection{Motivation}

The \Cyclus paradigm enables targeted contribution and collaboration within the
nuclear fuel cycle analysis community to achieve two important goals: lower the
barrier for users to include custom nuclear technologies and facility types in 
their fuel cycle
analyses while improving the ability to compare simulations with and without
those custom concepts.  This essential capability is absent in
previous simulators where user customization and extensibility were not design
objectives.  While the \emph{modular and open architecture} of
\Cyclus is necessary to meet these goals, it is not sufficient.
\emph{Agent interchangeability} is also required to facilitate direct comparison
of alternative modeling methodologies and facility concepts. With this concept
at its core, \Cyclus provides a platform for users to quickly develop the
capabilities at a level of detail and validation necessary for their
unique applications.  Finally, \Cyclus is applicable to a broader range of
fidelities, scales, and applications than other simulators, due to the
flexibility and generality of its \emph{\gls{ABM}} paradigm and \emph{discrete,
object-oriented approach}.

This structure recognizes that specialists should utilize their time and
resources in modeling the specific process associated with their area of
expertise (e.g., reprocessing and advanced fuel fabrication), without having to
create a model of the entire fuel cycle to serve as its host.  \Cyclus supports
them by separating the problem of modeling a physics-dependent supply chain into
two distinct components: a simulation kernel and archetypes that interact with
it. The kernel is responsible for supporting the deployment and
interaction logic of entities in the simulation.  Physics calculations and
customized behaviors of those entities are implemented within \emph{archetype}
classes.

Ultimately, modeling the evolution of a physics-dependent, international
nuclear fuel supply chain is a multi-scale problem which existing tools cannot
support. They have either focused on macro effects, e.g., the fleet-level
stocks and flows of commodities, or micro effects, e.g., the used-fuel
composition of fast reactor fuel. Each focus has driven the development of
specialized tools, rendering the task of answering questions between the macro
and micro levels challenging within a single tool.  In contrast, the open, extensible
architecture and discrete object tracking of \Cyclus allow the creation and
interchangeability of custom archetypes at any level of fidelity and by any
fuel cycle analyst.

\subsubsection{Open Access and Development Practices}

The proprietary concerns of research institutions and security constraints of
data within fuel cycle simulators often restrict access. Use of a simulator is
therefore often limited to its institution of origin, necessitating effort
duplication at other institutions and thereby squandering broader human
resources. License agreements and institutional approval are required for most
current simulators (e.g. \gls{COSI}6, \gls{DANESS}, \gls{DESAE}, EVOLCODE,
FAMILY21, \gls{NFCSim})\cite{juchau_modeling_2010}, including ORION,
and \gls{VISION}.  Even when the source code is unrestricted, the platform on which it relies (e.g. VENSIM) 
is often restricted or costly. The MIT \gls{CAFCA} software, for example, 
relies on the commercially licensed VENSIM product as a platform.
\Cyclus, on the other hand, is written in C++ for which freely available 
development tools and an open standard are available. Further, \Cyclus relies 
only on open source, freely available libraries. As such, it provides fully 
free and open access to all users and developers, foreign and domestic.

Moreover, both technical and institutional aspects of the software development
practices employed by the \Cyclus community facilitate collaboration.
Technically, \Cyclus employs a set of tools commonly used collaborative
software development that reduce the effort required to comment on, test and
ultimately merge individual contributions into the main development path.
For many of the simulation platforms adopted by previous simulators, there were
technical obstacles that impeded this kind of collaboration.
Institutionally, \Cyclus invites all participants to propose, discuss and
provide input to the final decision making for all important changes.

\subsubsection{Modularity and Extensibility}

Modularity is a key enabler of extending the scope of fuel cycle analysis
within the \Cyclus framework.  Changes that are required to improve the
fidelity of modeling a particular agent, or to introduce entirely new agents,
are narrowly confined and place no new requirements on the \Cyclus kernel.
Furthermore, there are few assumptions or heuristics that would otherwise
restrict the algorithmic complexity that can be used to model the behavior of
such agents.

For example, most current simulators describe a finite set of acceptable cycle
constructions (once through, single-pass, multi-pass). That limits the
capability to create novel material flows and economic scenarios. The \Cyclus
simulation logic relies on a market paradigm, parameterized by the user, which
flexibly simulates dynamic responses to pricing, availability, and other
institutional preferences.

This minimal set of mutual dependencies between the kernel and the agents is
expressed through the \gls{DRE} that provides a level of flexibility that does
not exist in other fuel cycle simulators.  It creates the potential for novel
agent archetypes to interact with existing archetypes as they enter and leave
the simulation over time and seek to trade materials whose specific
composition may not be known \textit{a priori}.

\subsubsection{Discrete Facilities and Materials}

Many fuel cycle phenomena have aggregate system-level effects which can only be
captured by discrete material tracking \cite{huff_next_2010}.  \Cyclus
tracks materials as discrete objects. Some current fuel cycle simulation tools
such as \gls{COSI}
\cite{mccarthy_benchmark_2012,grasso_nea-wpfc/fcts_2009,guerin_benchmark_2009},
FAMILY21\cite{mccarthy_benchmark_2012},
\gls{GENIUS} version 1, \gls{GENIUS} version 2, \gls{NFCSim}, and ORION also
possess the ability to model discrete materials. However, even among these, the ability to model reactor facilities individually is not equivalent to the ability to model distinct activities. \gls{COSI}, for example,
has some support for modeling reactors individually, but according to a recent
benchmark \cite{boucher_benchmark_2012}, it models many reactors operating in sync. That is, refuelling and discharging occur simultaneously for all reactors.
While \Cyclus allows this type of fleet-based aggregation of reactor behavior, \Cyclus also enables operations in each facility to vary independently of any others in the simulation.

Similarly, the ability to model disruptions (i.e. facility shutdowns due to
insufficient feed material or insufficient processing and storage capacity) is
most readily captured by software capable of tracking the operations status of
discrete facilities \cite{huff_next_2010}.  Fleet-based models (e.g.
\gls{VISION}) are unable to capture this gracefully, since supply disruptions
are modeled as a reduction in the capacity of the whole fleet.  All of the
software capable of discrete materials have a notion of discrete facilities,
however not all handle disruption in the same manner. \gls{DESAE}, for example,
does not allow shutdown due to insufficient feedstock. In the event of
insufficient fissile material during reprocessing, \gls{DESAE} borrows material
from storage, leaving a negative value \cite{mccarthy_benchmark_2012}.  The
\Cyclus framework does not dictate such heuristics. Rather, it provides a
flexible framework on which either method is possible.

A final benefit of the discreteness of facilities and materials is their power
when combined. The ability to track a material's history as it moves from one
facility to another is unique to \Cyclus. While some current simulators track
materials in discrete quanta, they do not necessarily preserve the identity of each quantum as the
materials move around the fuel cycle. When coupled with \Cyclus' individual
facility modeling, this capacity becomes distinct from what other fuel cycle
simulators are able to do. So, while FAMILY21 and \gls{COSI} can identify
whether a batch being discharged from a reactor originated in \gls{MOX}
fabrication rather than fresh \gls{UOX} fabrication, \Cyclus can go further,
tracking which of the fresh batches contained material from a particular
discharged batch. By extension, \Cyclus can also report which individual
facilities the batch passed through and in which it originated. 
The ability to track a single material through the simulation, 
though it might be split, transmuted, or merged with other materials along the 
way, allows \Cyclus to answer more data-rich questions that previous simulators
have been unable to ask. For example, it allows precise tracking of 
specific material diversions, so queries about nonproliferation 
robustness in a facility can be levied either in the context of a single event or 
a series of nefarious acts.

\section{Methodology and Implementation}

A modular, \acrfull{ABM} approach is ideal for modeling the coupled,
physics-dependent supply chain problems involving material routing, facility
deployment, and regional and institutional hierarchies which arise in fuel 
cycle analysis. Additionally, the choice
to build \Cyclus on open source libraries in modern programming languages
enables both remote and multiprocess execution on a number of platforms. This
section begins by describing the general design features that make \Cyclus both
flexible and powerful: \emph{cluster-ready software} and \emph{dynamically
loadable libraries}.  The
\gls{ABM} framework is then described, focusing on its implementation and benefits
in a fuel cycle context. A discussion of the
time-dependent treatment of discrete resources follows, focusing on the
\gls{DRE}. Support for users and developers via the \Cycamore library of
archetypes and the experimental toolkit are also presented.  Lastly, the methods
for quality assurance are outlined.

\subsection{Modular Software Architecture}

The architecture of \Cyclus allows developers to define nuclear fuel cycle
processes independent of the simulation logic. To achieve this, \emph{agents} are
developed which represent facilities, institutions, and regions comprising the
nuclear fuel cycle. These agents are created using the \Cyclus framework
\gls{API}, a set of functions and protocols which assist in agent development
and specify how agents should be defined.  This encapsulated `plug-in' design
choice provides two major benefits. First, analysts can take advantage of the
simulation logic \gls{API} and archetype ecosystem when they apply \Cyclus to
their specific problem.  A modeler can focus on creating or customizing
nuclear facility, institution, resource, and toolkit models within their
specific area of technical expertise. Second, because \Cyclus uses a modular
archetype approach, comparing two archetypes is straightforward. For example, if
an analyst would like to compare the effect of using different models to
determine the input fuel composition for fast reactors, fuel
fabrication archetypes can be developed and interchanged while keeping the rest
of the models used in the simulation fixed.

\subsubsection{Cluster-Ready Software}

Many fuel cycle simulators rely on \gls{COTS} and Windows-only software that 
limits performance on and compatibility with resource computing infrastructures 
(e.g.  cluster or cloud computing). This constrains the possible scope of 
simulations and increases the wall-clock time necessary to conduct 
parameterized sensitivity analyses and other multi-simulation studies.  For 
example, large scale sensitivity analyses to quantify the dependence of fuel 
cycle outcomes are only feasible in a massively parallelized environment.  
To enable such research, \Cyclus, is primarily written in \texttt{C++} and 
relies on libraries supported by Linux and UNIX (including Ubuntu and OSX) 
platforms, which are flexible and support parallelization.

Cyclopts \cite{gidden_cyclopts_2015}, a proof-of-principle \Cyclus-enabled 
application on such a large compute system,
uses UW-Madison's HTCondor \gls{HTC} infrastructure to study \gls{DRE} performance and
outcomes. For example, an investigation of the effect of solution degeneracy, a
commonly observed phenomenon in scenarios with individual facilities and basic
(e.g., commodity-only) preference definitions, was performed for three different
fuel cycles: once-through, \gls{MOX} fast reactor-thermal reactor cycles, and
\gls{MOX}-\gls{ThOX} recycle in fast reactors-thermal reactor cycles. Objective
coefficients were generated based on two factors: commodity-facility pairings
and facility location. The population of the possible values of
commodity-facility pairings is always small ($\mathcal{O}(10)$). The size of the
population of possible values of the location effect was investigated for values
of zero, ten, and infinity (i.e., any real number). The study additionally
included an investigation of problem-scaling behavior in order to quantify the
magnitude and rate-of-increase of \gls{DRE} solution times as a function of the
simulation-entity population for each fuel
cycle\cite{gidden_agent-based_2015}. In total, Cyclopts has run over $10^5$
jobs, comprising more than 60,000 compute hours. The \gls{HTC} infrastructure
has separately been utilized to run and collect information from full \Cyclus
simulations running in parallel on $10^3$ machines reliably for order $10^5$
simulations.

\subsubsection{Dynamically Loadable Libraries}

The \Cyclus architecture encourages efficient, targeted contribution to the ecosystem of
archetype libraries.
With \Cyclus, a researcher can focus on generating an archetype model within their
sphere of expertise while relying on the contributions of others to fill
in the other technologies in the simulation.
Similarly, individual developers may explore different levels of complexity within their archetypes, including
wrapping other simulation tools as loadable libraries within \Cyclus.

\begin{figure}[htbp]
\begin{center}
\includegraphics[width=\columnwidth]{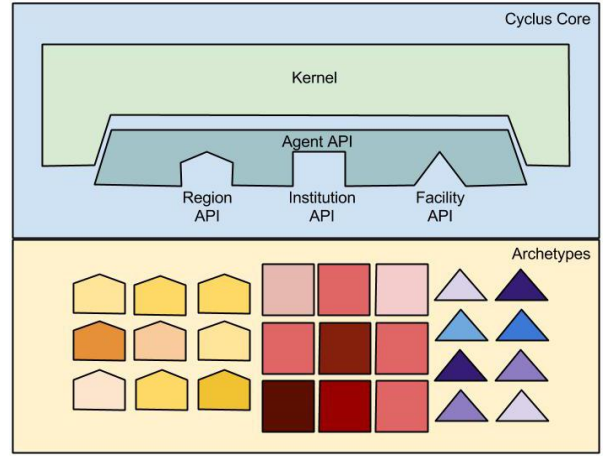}
\caption{The \Cyclus core provides \glspl{API} that abstract away the details in
the kernel and allow the archetypes to be loaded into the simulation in a modular
fashion.}
\end{center}
\label{fig:framework}
\end{figure}

\Cyclus achieves this behavior by implementing generic \glspl{API} and a
modular architecture via a suite of dynamically loadable plug-in libraries
(pictured in Figure \ref{fig:framework}). By anticipating the possible classes of
information required by the simulation kernel, the \Cyclus \glspl{API}
facilitate information passing between the plug-in agents and the core
framework.
Though common in modern software architecture, such a plug-in paradigm has not
previously been implemented in a nuclear fuel cycle simulator.
It allows the core \Cyclus framework to operate independently from the plug-in libraries, and the
dynamically loadable plug-ins to be the primary mechanism for extending \Cyclus'
capabilities independent of the core.

\begin{figure}[htbp]
\begin{center}
\includegraphics[width=0.8\columnwidth]{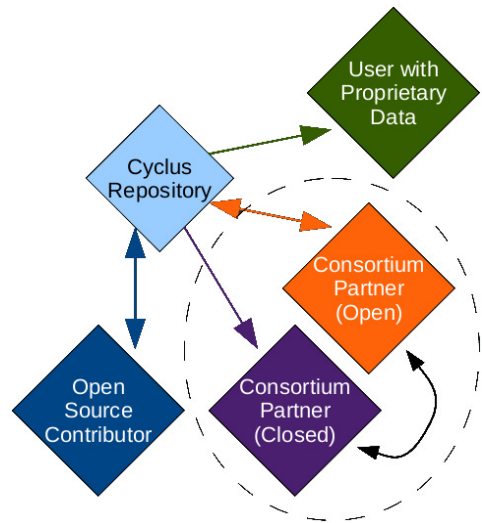}
\caption{The \Cyclus framework enables fully open, partially open, and fully
closed collaborations\cite{carlsen_cyclus_2014}.}
\end{center}
\label{fig:modifiedopen}
\end{figure}

An additional benefit is the ability for
contributors to choose different distribution and licensing strategies
for their contributions. Users and modelers control the accessibility of their archetypes and data sets (See Figure \ref{fig:modifiedopen}).
In particular, since the clean plug-in architecture loads libraries without any
modifications to the \Cyclus kernel, closed-source archetypes can be used with
the simulator alongside open source archetypes without transfer of sensitive information. This architecture
allows closed-source libraries (e.g., those representing sensitive nuclear
processes and subject to export control) to be developed and licensed privately.

Finally, dynamically loadable libraries enable \Cyclus to easily handle varying levels of simulation complexity. Hence a single
simulation engine can be used by both users interested in big-picture policy
questions as well as users focused on detailed technical
analyses. They merely choose their preferred level of modeling depth from among the
available libraries in the ecosystem.

\subsection{Agent-based Paradigm}
\label{sec:abm}

\Cyclus implements an \acrlong{ABM} paradigm. \gls{ABM} enables model
development to take place at an agent level rather than a system level. In the
nuclear fuel cycle context, for example, an analyst can design a reactor agent
that is entirely independent from a fuel fabrication agent. Defining the
behavior of both agents according to the
\gls{API} contract is sufficient for them to interact with one another as
bona fide agents in the simulation.  The two archetype libraries can be used in
the same simulation without any shared knowledge, allowing modelers to
construct a simulation from building blocks of many types and origins.

Furthermore, the \gls{ABM} paradigm is more flexible and intuitive (from both a developer and user perspective) than the system dynamic approach used in current simulators.
System dynamics is a popular approach for modeling nuclear fuel cycles
\cite{jacobson_vision_2009,van_den_durpel_daness_2009,guerin_impact_2009,guerin_benchmark_2009}.
Formally however, system dynamics models are simply a strict subset of agent-based models
\cite{macal_agent-based_2010}.
That is, any system dynamics model can be translated
into an agent-based model.
\gls{ABM} techniques therefore enable a broader range of simulations in a more
generic fashion.

\subsubsection{Agent Interchangeability}\label{sec:interchangeability}


\gls{ABM} is inherently object-oriented because agents represent discrete,
independently acting objects.  Figure \ref{fig:framework} illustrates the
modular nature of \Cyclus archetypes.  The core of the \Cyclus simulator creates
a set of classes on which agent plug-ins are based.  Agent plug-ins utilize the
generic core \glspl{API} that define agent-to-agent interaction as well as agent-to-environment interaction.
For example, they use the
resource exchange paradigm \gls{API} for trading resources with one another.  
For the archetype developer, these interfaces provide enormous power
simply. The \gls{API} abstracts away details unnecessary to specifying the 
archetype behavior, while providing
all necessary functionality for interacting with the \Cyclus simulation kernel.
For the user, multiple archetypes that inherit the same \glspl{API} are interchangeable
in a simulation.

\begin{figure}[htbp]
\begin{center}
\includegraphics[width=\columnwidth]{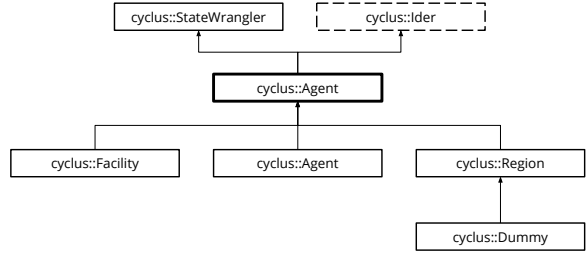}
\end{center}
\caption{Inheriting \Cyclus class interfaces, such as the \Class{Agent},
\Class{Facility}, \Class{Institution}, and \Class{Region} classes, abstracts away
unnecessary details while exposing powerful functionality. In the above
example, the \Class{Dummy} archetype simply inherits from \Class{Region} in
order to become a bona fide Region-type \Class{Agent}.}
\label{fig:agent_uml}
\end{figure}

Critically, this novel functionality enables the comparison
between agent implementations. For example, 
an archetype that inherits the \Class{Region} interface, as in Figure
\ref{fig:agent_uml}, is interchangeable with any other Region agent.

In this way, a researcher can directly compare two different reactor modeling
implementations (perhaps the imaginary classes \Class{DetailedReactor} and \Class{SimpleReactor})
simply by exchanging the two corresponding archetypes. That is, two reactor
archetypes both inheriting from the \Class{Facility} class are indistinguishable
from a simulation perspective.  This can be done with any agent type, where agents can be ``Regions,'' ``Institutions,'' or ``Facilities.''

\subsubsection{Regions, Institutions, and Facilities}
\label{sec:rif}

\Cyclus provides a novel representation of entities in the nuclear fuel cycle
that reflects the reality in international nuclear power: facilities implementing individual fuel cycle technologies, institutions managing those facilities, and regions providing geographical and political context for institutions and facilities. While
few simulators have provided any notion of static regional
effects \cite{huff_next_2010,juchau_modeling_2010}, \Cyclus allows for both regions and institutions to be first-class
agents in simulated fuel cycles. The fundamental interactions for each entity are implemented in a corresponding
archetype class in \Cyclus, i.e., the \Class{Region} class, \Class{Institution}
class, and \Class{Facility} class. Archetype developers can then build on the
provided functionality by inheriting from the appropriate class.
\Cyclus implements a \gls{RIF} relationship through a
parent-child hierarchy where
regions are the parents of institutions which are, in turn, the parents of
facilities. In other words, \gls{RIF} hierarchies form a \gls{DAG}\footnote{
\gls{DAG}s are common graph theoretic structures whose most important feature is a 
lack of \emph{cycles}, i.e., there is a single path from the root node to any other 
node in the graph. 
},
with regions as root nodes and facilities as leaf nodes.

Two primary consequences arise from this structure. First, institutions are
nominally responsible for deploying and decommissioning facilities. Accordingly, advanced
logic regarding building and decommissioning can be implemented on top of
those behaviors inherited from the
\Class{Institution} interface. Second, the \Class{Facility} class implements the
\Class{Trader} interface to participate in resource exchange, and institutions and regions, respectively, can
adjust the resource flow preferences of their managed facilities. Importantly,
this capability allows for the modeling of preferential regional trading
of resources (e.g., tariffs) as well as preferential institutional trading
(e.g., long-term contracts).

\subsection{Discrete Objects}

\Cyclus models facilities, institutions, regions, and resources as discrete
objects. A discrete resource model allows for a range of modeling granularity. In the
macroscopic extreme, it is equivalent to time-stepped continuous flow. In the
microscopic extreme, the model is capable of representing arbitrarily small
material objects at isotopic resolution. In this way, \Cyclus is
applicable across the full range of modeling fidelity.

Fleet-based, lumped-material models do not distinguish between discrete facility
entities or materials. However, some questions require resolution at the level
of individual facilities and materials. As a result, many detailed performance
metrics cannot be captured with previously existing fleet-based models. For 
example, meaningful models of spent nuclear fuel storage transport, and
disposal strategies, require representation of discrete casks and their varying
isotopic compositions.

For all
of the reasons that the \gls{ABM} paradigm in \ref{sec:abm} enables novel
simulations, multiple use cases require that these agents, such as the regions,
institutions, and facilities in \Cyclus, must be represented as discrete
objects. For instance, tracking of individual shipments is only viable if materials and
resources are tracked as discrete objects.

\subsubsection{Resources and Materials}
\label{sec:mats}

Another such use case seeks to capture system vulnerability to
material diversion. Provenance and trade-history of distinct materials is the fundamental
information unit in such studies, and so this type of analysis requires
 discrete simulation of a
target facility and the individual materials modified within it.
Material risk analysis, therefore, demands that both facilities and materials
should be discretely modeled objects like those in \Cyclus.

In \Cyclus, agents can transfer discrete resource objects among one another.
Cyclus supports two types of resources:

\begin{itemize}

  \item materials: these represent typical nuclear materials with
      nuclide compositions;

  \item products: these can represent any user-defined measure: carbon
      credits, build permits, employees, etc..

\end{itemize}

All operations performed on every resource object (splitting, combining,
decay, etc.) are tracked in detail as they are performed.  This information
includes the agent that created each resource when it was introduced into the
simulation.  The parentage of each resource is also tracked. This makes it
possible to follow the history of resources as they are transferred between
agents.

The \Cyclus kernel has built-in experimental support for decay calculations.
Materials store the time since their last decay and agents are free to
invoke the decay function on them as desired to decay them to the current
simulation time. \Cyclus can operate in 3 decay modes, with 1 additional
mode likely to be added in a future release:

\begin{itemize}
    \item ``manual'' (currently implemented) is the default mode
        where agents decay materials when requested by an archetype,
    \item ``never'' (currently implemented) globally turns off all decay.
        The Material decay function does nothing,
    \item ``lazy'' (currently implemented) decays material automatically whenever
         its composition is observed (e.g. when an agent queries information
         about a material's $^{239}$Pu content),
    \item ``periodic'' (future) automatically decays all materials in a
        simulation with some fixed frequency.
\end{itemize}

When decay is invoked, a material checks to see if it contains any nuclides with
decay constants that are significant with respect to the time change since the
last decay operation.  If none of the decay constants are significant, no decay
calculation is performed and the material remains unchanged.  This error does
not accumulate because the next time the material's decay function is invoked,
the time change will be larger.

\Cyclus has no notion of ``tracked'' versus ``untracked'' nuclides.  In
\Cyclus, the composition of a material is represented by an arbitrarily large
list (potentially thousands) of nuclides.  Agents are free to treat nuclides
present in materials any way they please - including ignoring them.  It is the
responsibility of archetype developers to choose how to handle potentially
full-fidelity compositions.

\begin{figure}[htbp]
\begin{center}
\includegraphics[width=\columnwidth]{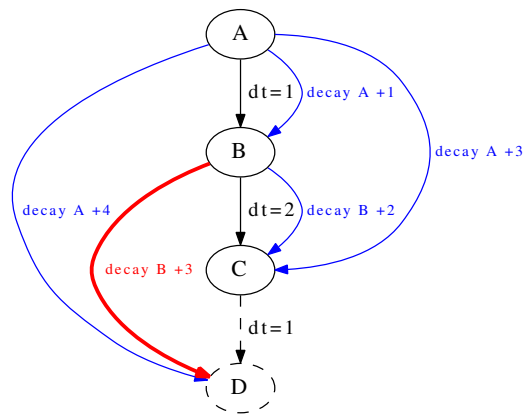}
\end{center}
\caption{
A simple decay line cache holding compositions A, B, C, and a
yet-uncomputed composition D.  B comes from decaying A 1 time step.  C comes
from decaying B 2 time steps, etc.  Black represents the cache for this
particular composition chain. Blue indicates decay operations that can be
satisfied by cache lookups.   If A needs to be decayed 1 time step (A +1), a
quick lookup returns the previously computed B.  Decaying B 3 time steps
requires a decay calculation to compute a new composition D, but subsequent
requests such as decaying A 4 time steps will not require any recalculation.
}
\label{fig:compositions}
\end{figure}

In large simulations, many material objects may change frequently.  Material
decay can also contribute significantly to such changes.  In order to help
avoid unnecessary runtime performance and database size impacts, compositions
in \Cyclus have some special features.  In particular, compositions are
immutable once created. This allows multiple material objects to hold
references to the same composition safely.  Additionally, new compositions
resulting from decay are cached and used to avoid redundant decay
calculations.  Figure \ref{fig:compositions} illustrates how this decay
history cache works. Composition immutability in concert with decay history
caching help eliminate many redundant calculations in addition to reducing the
total number of composition entries recorded in the database.

\subsubsection{Dynamic Resource Exchange (DRE)}

The \Cyclus simulation paradigm allows discrete agents, based on archetypes
about which the kernel has no knowledge, to enter the simulation at arbitrary
times and trade in discrete resources. These resources are not defined \textit{a
  priori}. Therefore, the logic engine defining resource interaction mechanisms
among agents is crucial. The \gls{DRE}, described in detail in
\cite{gidden_agent-based_2015}, is that critical logic engine on which \Cyclus
simulations are built.  Supporting the general \Cyclus philosophy, facilities
are treated as black boxes and a supply-demand communication framework is
defined.

The \gls{DRE} consists of three steps: supply-demand information
gathering, resource exchange solution, and trade execution. Importantly, each
step is agnostic with respect to the exchange of resources in question, i.e.,
the same procedure is used for both Materials and Products.

The information-gathering step begins by polling potential consumers. Agents
define both the quantity of a commodity they need to consume as well as the
target isotopics, or quality, by posting their demand to the market exchange as
a series of \textit{requests}. Users may optionally parameterize the agent to
associate a collection of demand constraints with each collection of requests.
Collections of requests may be grouped together, denoting \textit{mutual}
requests that represent demand for
some common purpose. For example, a reactor may request \gls{UOX} and \gls{MOX} fuel
mutually, indicating that either will satisfy its demand for fuel.

Suppliers then respond to the series of requests with a \textit{bid}. A bid
supplies a notion of the quantity and quality of a resource to match a
request. Suppliers may add an arbitrary number of constraints to accompany
bids. For example, an enriched \gls{UOX} supplier may be constrained by its current
inventory of natural uranium or its total capacity to provide enrichment
in \glspl{SWU}. It attaches such constraints to its bids.

Any potential resource transfer, i.e., a bid or a request, may be denoted as
\textit{exclusive}. An exclusive transfer excludes partial fulfilment; it must
either be met fully or not at all. This mode supports concepts such as the
trading of individual reactor assemblies.  In combination with the notion of
mutual requests, complex instances
of supply and demand are enabled.

Finally, requesting facilities, institutions and regions may apply
\textit{preferences} to each potential request-bid pairing based on the proposed
resource transfer. Facilities can apply arbitrary complex logic to rank the bids
that they have received, whether based on the quantity available in each bid or
on the quality of each bid, and the consequent implications of the physics behavior
of that facility. In addition, an institution can apply a higher preference to a
partner to which it is congenial; similarly, a region may negate any transfers
of material which have a higher uranium enrichment than is allowable.

\begin{figure}[htbp]
\begin{center}
\includegraphics[width=\columnwidth]{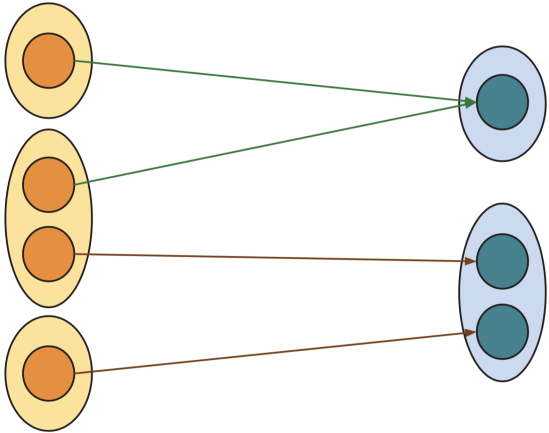}
\end{center}
\caption{The flow graph representing three suppliers (left), two requesters (right), 
and the potential flows of various commodities among them. 
The second consumer makes two different requests.
Meanwhile, the second supplier can supply the commodities 
requested by both consumers and has provided two bids accordingly 
\cite{gidden_agent-based_2015}.}
\label{fig:bid_req}
\end{figure}

Given a full definition of supply and demand, represented in Figure 
\ref{fig:bid_req} as a flow graph, the \gls{DRE} may be solved either
optimally using a mathematical program or approximately by a simulation-based
heuristic \cite{gidden_agent-based_2015}. If any trade is denoted as exclusive, e.g., if an
analyst desires an assembly-fidelity model, then either a heuristic must be used
or exchanges must be represented as a \gls{MILP}. If no exclusive trades exist,
a \gls{LP} model may be used. The \gls{LP} formulation in \Cyclus is of the 
form given by Gidden \cite{gidden_agent-based_2015}:

\begin{align}
\min_{x} \: \: & z = c^\top x \\
\text{s.t.} \: \:  & A x \leq b \\
    & x_{i,j} \in [0, \tilde{x}_j] \: \:  \forall i \in I, \forall j \in J
\end{align}
\begin{align}
\intertext{where}
I &= \mbox{set of supply nodes}\nonumber\\
J &= \mbox{set of request nodes}\nonumber\\
c_{i,j} &= \mbox{unit cost of flow from i to j}\nonumber\\
x_{i,j} &= \mbox{flow from node i to node j}\nonumber\\
a^k_{i,j} &= \mbox{coefficient for constraint k between i and j}\nonumber\\
b^k_{s \mid r} &= \mbox{RHS for constraint k for a given supplier or requester}\nonumber\\
\tilde{x}_j &= \mbox{requested quantity of request node j.}\nonumber
\end{align}

In practice, \glspl{LP} solve much faster than
\glspl{MILP}. However, both capabilities exist in \Cyclus in order to
provide users with the desired level of fidelity.

Trades between agents are initiated by the \Cyclus kernel after a solution to
the \gls{DRE} is found. For each trade, the supplying agent is notified of its
matched request and provides an associated resource to the exchange. All
supplied resources are then sent to the corresponding requesting agents.

In \Cyclus, the \gls{DRE} is executed at each time step. Therefore, if a
facility's request for a resource is not met at a given time step, it may offer
a request in the following time step. Because agent behavior may change
arbitrarily, the exchange executed at any given time step may be unique in a
simulation.

The \gls{DRE} is a novel simulation concept in the nuclear fuel cycle domain. It
provides a flexible supply-demand infrastructure, supporting dynamic flows of
resources between agents, even as those agents enter and leave the simulation, and
even when those agents are defined by archetypes of arbitrary complexity. Trading
between agents can be affected by both the
proposed quality of a resource and agent relationships through the use of
preferences. Accordingly, a wide range of possible effects can be
modeled, from capacity-limited fuel supply to international trade agreements.

\subsection{Simulation Support}
So that users and developers can build working simulations in the shortest time
possible, the \Cyclus ecosystem provides fundamental building blocks: basic
archetypes and a toolkit of commonly needed functions.  The \Cycamore library
provides a suite of fundamental Region, Institution, and Facility archetypes,
while the \Cyclus toolkit provides assistance to developers.

\subsubsection{Cycamore}

\Cycamore \cite{carlsen_cycamore_2014}, the \Cyclus additional module
repository, provides a fundamental set of dynamically loadable libraries
providing agent archetypes for basic simulation
functionality within \Cyclus.  Since \Cyclus relies on external
archetypes to represent the agents within a simulation, \Cycamore provides the
basic archetypes a new user needs to get started running simple simulations.
These archetypes support a minimal set of fuel cycle simulation goals and
provide, by example, a guide to new developers who would seek to contribute
their own archetypes outside of \Cycamore.

As of version 1.0, \Cycamore contains one region archetype, two institution
archetypes, and four facility archetypes. Three additional facilities were
added in version 1.3 . Short descriptions of these functions can be found in
Table \ref{tab:cycamore}.

\begin{table*}[htb]
\centering
\begin{tabularx}{\textwidth}{rlX}
\hline
\textbf{Entity} & \textbf{Archetype} & \textbf{Functionality} \\
\hline
Facility & EnrichmentFacility & This facility enriches uranium at a specified capacity. \\
Facility & Fab$^*$ & This facility fabricates fuel material based on separated streams. \\
Facility & Reactor$^*$ & A reactor model that handles batch refueling, based on pre-determined recipes of compositions. It requests any number of input fuel types and transmutes them to static compositions upon discharge.\\
Facility & Separations$^*$ & This facility separates an incoming material into specified streams. \\
Facility & Sink & This facility is capable of accepting a finite or infinite quantity of some commodity produced in the simulation. \\
Facility & Source & This facility generates material of the composition and commodity type specified as input.  \\
Institution & ManagerInst & The manager institution manages production of commodities among its facilities by building new ones as needed. \\
Institution & DeployInst &  This institution deploys specific facilities as defined explicitly in the input file. \\
Region & GrowthRegion & This region determines whether there is a need to meet a certain capacity (as defined via input) at each time step. \\
\hline
\end{tabularx}
\caption{The Archetypes in \Cycamore seek to cover a large range of simple
simulation use cases \cite{carlsen_cycamore_2014}. Facilities added in version
1.3 are marked with $^*$.}
\label{tab:cycamore}
\end{table*}

Section \ref{sec:simulations} will demonstrate how the current \Cycamore
release provides basic functionality enabling simple fuel cycle analyses. As
future contributions are vetted, the capabilities in \Cycamore may grow.

\subsubsection{Toolkit}

In addition to the core functionality of the \Cyclus kernel, which is focused on
the set of capabilities needed to implement an agent-based simulation
with \gls{DRE}, a toolkit is provided to assist developers
and users with related simulation and nuclear engineering tasks. The toolkit is
an actively developed part of \Cyclus, with a primarily forward-looking
focus on supporting interesting \textit{in situ} metric analysis tools.

\paragraph{Simulation Tools}

A series of utility classes are provided to support demand-constrained agent
facility deployment. For example, symbolic function representations of linear,
exponential, and piecewise functions are supported via the
\Class{SymbFunctionFactory} class. Such functions are used with other toolkit
classes to determine commodity demand (e.g., power demand) from user input. Four
mix-in classes provide the basis for in-simulation deployment determination:
\Class{CommodityProducer}, \Class{CommodityProducerManager}, \Class{Builder},
\Class{BuildingManager}. The \Class{CommodityProducer} class provides an
interface for querying the \textit{prototypes} which have the
capacity to produce a given commodity. The \Class{CommodityProducerManager}
provides an interface for registering \Class{CommodityProducer}s and querying
the current capacity (supply) of a commodity. The \Class{Builder} class provides
an interface for querying which prototypes can be built and interacts with the
\Class{BuildingManager}, which orders prototypes to be built. The
\Class{BuildingManager} uses a simple minimum cost algorithm to determine how
many of each prototype, $y_i$, to build given a demand ($\Phi$), capacities
($\phi_i$), and costs ($c_i$). Here $i$ indexes $I$ available prototypes which perform a similar function, and the demand, capacity and cost carry prototype-specific units which are defined by the developer.

\begin{equation}
\begin{aligned}
 \min & \sum_{i=1}^{N}c_i y_i \\
 s.t. & \sum_{i=1}^{N}\phi_i y_i \ge \Phi \\
      & y_i \in [0,\infty) \:\: \forall i \in I, \:\: y_i \:\: \text{integer}
\end{aligned}
\end{equation}

\paragraph{Nuclear Engineering Tools}

The \Cyclus toolkit provides two useful interfaces for querying physical parameters of \Class{Material}
objects. First, the \Class{MatQuery} tool
provides a basic querying \gls{API}, including the atom and mass fractions of
nuclides, the number of moles of a nuclide in a material, etc. (i.e., a
\Class{Composition}), in a material. The
\Class{Enrichment} tool provides an \gls{API} for determining enrichment-related
parameters of a material, including the \gls{SWU} and natural
uranium required to enrich a material provided knowledge of feed, product, and tails
assays.

\paragraph{Toolkit Extensions}

In addition to those that already exist, new tools will
emerge from the archetypes developed by the community. As these tools gain adoption between projects and demonstrate their
utility to the developer community, they will be considered for screening and
adoption into the kernel as toolkit extensions. Likely extensions include

\begin{itemize}
\item fuel cycle metrics calculators,
\item supportive data tables,
\item policy models,
\item and economic models.
\end{itemize}

\subsection{Quality Assurance}


To garner the trust of a broad user and archetype developer community, the
\Cyclus project must implement a strategy to assure the ongoing quality of the
software it provides.  Multiple strategies, collectively known as
\emph{\gls{QA}}, have been developed by the scientific software development
community to mitigate structural and algorithmic errors in software. These
include \emph{\gls{VV}} \cite{boehm_software_1989}, testing, and others.

Nuclear engineering software quality is often governed by \gls{NQA1}, an
\gls{ASME} specification whose latest revision appeared in 2009
\cite{asme_nqa-1a-2009_2009}.  \Cyclus has adopted an \emph{agile} development
process \cite{larman_agile_2004}, interpreting \gls{NQA1} in a manner similar
to the process adopted by the \gls{DOE} within \gls{NEAMS}
\cite{neams_nuclear_2013} or by the PyNE toolkit \cite{biondo_quality_2014}.

Validation of simulators like \Cyclus, that are intended to give forecasts of
system behavior in uncertain futures, is not well defined as there are no
experimental benchmarks upon which to rely.  Instead, code-to-code comparisons
with fuel cycle simulators that use other modeling paradigms are underway as
the best approach to establish confidence that \Cyclus produces correct
answers. \cite{huff_extensions_2014}

Verficiation of \Cyclus implementation relies on software development best
practices such as version control, testing, continuous integration,
documentation, and style guidelines to ensure reliability and
reproducibility. Sections \ref{sec:qa-vc}-\ref{sec:qa-review} discuss in
greater detail the software development components that comprise the \Cyclus
verification strategy, which allows the simulation kernel and physics models
to be tested explicitly and separately. Each of these approaches on its own is
a valuable addition to \gls{QA} but it cannot be the entire answer to the
requirements imposed by \gls{NQA1}. Taken together and strictly adhered to,
they present a fortress to protect against poorly designed or otherwise
undesirable code.

\subsubsection{Version Control}
\label{sec:qa-vc}

Automated version control, provided by git, a well-established \emph{distributed
  version control} tool \cite{software_freedom_conservancy_git_2014}, is
at the heart of the QA strategy because it records the full history of any
change, along with metadata such as the author, a timestamp, and an
accompanying message. This makes it possible to identify when changes were
introduced, how they are related to other changes, and who made those changes.  If
necessary, it also facilitates reversing individual changes from a long and
intricate set of changes.  Version control also enables a code review process
descried in section \ref{sec:qa-review} below.

\subsubsection{Testing}
\label{sec:qa-testing}

Automated software \emph{testing} is the first line of defense against
errors in implementation. Testing directly compares the
actual results of running software versus the expected behavior of the software.
In \Cyclus, three categories of tests are defined: unit tests, integration
tests, and regression tests.  Before a proposed code change is allowed into 
\Cyclus,  the change must be covered by a test, either new or existing, and all 
tests must pass.

\paragraph{Unit Tests}

Unit tests verify behavior of the smallest code \emph{unit}, typically a
single function or a class.  \Cyclus uses the Google Test framework
\cite{inc_googletest_2008} as a harness for running unit tests. Sufficient unit
tests are required for any proposed change to the \Cyclus code base. Currently,
\Cyclus implements over 450 unit tests and \Cycamore implements 85.  These
cover approximately 65\% of their respective code bases, and these numbers are
expected to grow over time.

\paragraph{Integration Tests}

Integration tests combine multiple elements of the
\Cyclus interface and test that they work correctly with each other. 
In \Cyclus and \Cycamore, integration tests are performed by running sample
simulations for scenarios verifying that results match predictions. A set of standard input
files are run, then the output is inspected and compared via Nose
\cite{pellerin_nose_2007}, a Python test framework.  

\paragraph{Regression Tests}

Regression tests ensure significant unintended changes do not
occur over the course of \Cyclus development.
Regression tests are implemented similarly to integration tests.
In this category, however, the comparison is done against
the output of the same input file when run with a previous version of \Cyclus,
typically the last released version.
In some sense, regression tests are `dumb' in that they do
not care about the contents of a simulation being correct, only whether or not
it changed.

\paragraph{Continuous Integration}

\Gls{CI} is the idea that software should be automatically tested
and validated as it is being developed, rather than as a final stage in a
longer development cycle.  The results of this
testing are shown to code reviewers prior to reviewing the software changes themselves.  The \Cyclus
project uses the CircleCI \cite{biggar_circleci_2015} service for continuous integration.
When a code change is submitted for review, CircleCI builds a version of the \Cyclus
source code that includes the requested changes and runs the complete test
suite, reporting back whether or not those steps were successful. If \gls{CI}
was not successful, the code author(s) must first identify and resolve the 
problems.  These steps are performed for all incoming code prior to inclusion, 
so broken code never enters the main software development branch.

\subsubsection{Code Review}
\label{sec:qa-review}

Automated testing including \gls{CI} is a necessary but not sufficient
component of the \Cyclus \gls{QA} system: it keeps bad code out of
\Cyclus. However, \Cyclus will always require human eyes and hands to let good
code in.  

The main \Cyclus repository is hosted remotely and publicly on the GitHub
website \cite{dabbish_social_2012}, in part because it provides tools that
facilitate a culture of frequent and thorough code review.  A number of
policies exist to ensure that a proposed set of changes, known as a \emph{pull
  request}, adhere to the projects \gls{QA} standards. Every pull request must
be reviewed and accepted by a member of the \Cyclus core team that was not
involved in authoring the changes.  In addition to reviewing the algoirthmic
design in the source code, the reviewer relies on tests to ensure correct
behavior, and requires the authors to adhere to the style guide and provide
sufficient documentation.  Only after the \gls{QA} standards have been met are
the proposed changes merged into the software. This step has been repeatedly
shown to improve code quality \cite{cohen_modern_2010}.

Once \gls{CI} is successful, a code reviewer inspects not only the changes that 
are proposed to the software itself, but also the changes that have been 
proposed to tests.  

Because of their long term benefit to the maintainability of the project, 
documentation and coding style are also reviewed as part of this process.  The 
\gls{API} must be documented as required by the \Cyclus \gls{QA} policy.  In 
\Cyclus, this information is aggregated together into static websites with the 
Doxygen \cite{van_heesch_doxygen:_2008} and Sphinx \cite{brandl_sphinx_2014} 
tools, and can be accessed at \url{http://fuelcycle.org}. \Cyclus also strictly 
enforces the use of the Google C++ Style Guide \cite{weinberger_google_2008} 
for all software contributions.  This means that all developers of \Cyclus 
write \Cyclus code in the same way.  This homogenization may be a hurdle to new 
developers but ultimately improves code legibility and, therefore, robustness 
\cite{cohen_modern_2010}.

\section{Demonstration}
%

%


The success of the \Cyclus simulator can be measured in many ways.  The most
compelling are early successes of the community-driven development model and 
demonstrations of its fundamental simulation capabilities. Promising
growth of the \Cyclus ecosystem at multiple institutions indicates that a fuel
cycle simulator can advance in this community-driven development
paradigm. Additionally, simulation results for both once-through and more
complex recycling scenarios demonstrate that \Cyclus possesses the fundamental
fuel cycle simulation capabilities to contribute to the field.

\subsection{Ecosystem}

The \Cyclus community-driven software development model seeks to break the
proliferation of specialized simulators. It instead leverages
the collective expertise of fuel cycle researchers toward a single, more
extensible, tool. Through the targeted
contributions of those researchers, an ecosystem of capabilities should emerge.
The \Cyclus `Ecosystem' is the collection of tools, calculation libraries,
archetypes, data, and input files intended for use with the \Cyclus simulator.
Members of the ecosystem include:
\begin{itemize}
\item the archetypes provided in the \Cycamore \cite{carlsen_cycamore_2014}
repository
\item the archetypes created by researchers
\item isotopic composition data
\item historical facility deployment data
\item the \Cyclus \gls{GUI} tool Cyclist
\item fundamental analysis tools in the \Cyclus toolkit
\item tools for \Cyclus optimization, parallelization, and development
\end{itemize}
Taken together, these form an ecosystem of capabilities. Over time, this
ecosystem will grow as archetype developers, kernel developers, and
even users contribute capabilities developed for their own needs. Indeed, the
long-term vision for the \Cyclus framework predicts an ever-expanding ecosystem
of both general and specialized capability extensions.

Already, the ecosystem is growing. Early cross-institutional contributions to
the ecosystem demonstrate a significant achievement by the \Cyclus framework
and provide the basis for a community-driven development model.

\subsubsection{Supplementary Projects}

A number of projects and tools outside of the core simulation kernel have been
developed to improve the scope and the diversity of the capabilities in the
\Cyclus ecosystem. Table \ref{tab:coretools} lists the tools and projects
developed under close integration with the \Cyclus kernel.  These tools are used
to ease development and simulation design (Cycstub, Cycic, Ciclus), data
visualization and analysis (Cyclist, Cyan), and remote execution (Cloudlus).

\begin{table*}[htb]
\centering
\begin{tabularx}{\textwidth}{rXr}
\hline
\textbf{Name} & \textbf{Description} & \textbf{Ref.} \\
\hline
Cycic &  Input control embedded in Cyclist & \cite{flanagan_input_2013}\\
Cyclist & Interactive data exploration environment & \cite{livnat_cyclist_2014} \\
Ciclus & Continuous integration scripts for \Cyclus & \cite{scopatz_ciclus_2014}\\
Cycstub & Skeleton for clean slate module development & \cite{carlsen_cycstub_2014}\\
Cyan & \Cyclus analysis tool & \cite{carlsen_cyan_2014}\\
Cloudlus & Tools for running \Cyclus in a cloud environment & \cite{carlsen_cloudlus_2014} \\
\hline
\end{tabularx}
\caption{Many tools have been developed outside of the scope of the \Cyclus kernel for improved user, developer, and analyst experiences with \Cyclus.}
\label{tab:coretools}
\end{table*}

\subsubsection{Archetype Contributions}

It is expected that the most common type of contribution to \Cyclus will be
contributions of new archetypes. Researchers will be driven to create a new
archetype when a need arises, such as to improve the fidelity of simulation,
or to represent a novel reactor type, an innovative
reprocessing strategy, or a particular governmental or institutional behavior.
The real-world utility of \Cyclus can in part be measured by the breadth and
diversity of archetypes being developed in this way.

Early progress has been promising. Many archetypes external to the \Cycamore
library (Table \ref{tab:archetypes}) have been
\cite{huff_streamblender_2014,huff_commodconverter_2014} or are being
\cite{flanagan_bright-lite_2014,skutnik_development_2015,huff_mktdriveninst_2014}
developed for contribution to the \Cyclus ecosystem. These archetypes provide
the first examples of developer-contributed capabilities.  They add to the
fundamental \Cycamore archetypes by providing physics-based reactors,
separations logic, fuel fabrication processing, storage facilities, and expanded
institutional paradigms.  The existence and diversity of these contributed
archetypes illustrate the power and potential of the community-based development
approach that \Cyclus has taken.

\begin{table*}[htb]
\centering
\begin{tabularx}{\textwidth}{rXr}
\hline
\textbf{Name} & \textbf{Description} & \textbf{Ref.} \\
\hline
Bright-lite & A physics-based reactor archetype and fuel fabrication archetype & \cite{flanagan_bright-lite_2014} \\
Nuclear Fuel Inventory Model & A flexible, ORIGEN-based, reactor analysis module & \cite{skutnik_development_2015} \\
CommodConverter & A simple commodity converting storage facility archetype  & \cite{huff_commodconverter_2014} \\
MktDrivenInst & An institution that controls deployment based on commodity availability & \cite{huff_mktdriveninst_2014} \\
SeparationsMatrix & A facility for elemental separations of used fuel & \cite{huff_streamblender_2014} \\
StreamBlender & A facility for fuel fabrication from reprocessed streams & \cite{huff_streamblender_2014} \\
\hline
\end{tabularx}
\caption{A diverse set of archetypes under development reflect the diverse
needs of researchers at various institutions. These archetypes, contributed
outside of the \Cyclus core and \Cycamore libraries are the first demonstration
of community-driven development in a fuel cycle simulator.}
\label{tab:archetypes}
\end{table*}

\subsection{Simulations}
\label{sec:simulations}


To illustrate the flexibility of \Cyclus, this section will discuss the 
creation and results of a range of representative fuel cycle simulations.
Previous benchmarks between 
\Cyclus and system dynamics simulators for more complex problems, including 
transition analyses, have been reported 
elsewhere and \Cyclus has demonstrated satisfactory performance
\cite{djokic_application_2015,scopatz_non-judgemental_2015,gidden_agent-based_2015,gidden_once-through_2012} . 
The examples presented here are not the limit of \Cyclus' capabilities, which are extended with each new
addition to the ecosystem, as discussed above.  Rather, these simulations are
designed to illustrate that \Cyclus matches the capabilities of any
recipe-based simulator and that variations on a single \Cyclus simulations can 
be run with small changes to the input specification. 
For simplicity of the current demonstration, many simplifying assumptions have been
made with respect to material compositions, fuel transmutation, among others.
The three fuel cycles examined are:

\begin{enumerate}
    \item No Recycle (once through)
    \item 1-pass \gls{MOX} Recycle
    \item Infinite-pass \gls{MOX} Recycle
\end{enumerate}

For each of these fuel cycles, a 1,100 month single-reactor \Cyclus simulation
was run in addition to a 10-reactor simulation with staggered refueling times.
As the number of staggered-cycle reactors increases the system converges
toward continuous material flow results.  \Cyclus flexibility allows this
transition to be examined.  

\begin{table*}[htb]
\centering
\begin{tabularx}{\textwidth}{XrX}
\hline
\textbf{Facility} &\textbf{Parameter} & \textbf{Value}\\
\hline
\textbf{LWR} & type & LWR\\
\Class{cycamore::Reactor} & batches & $3$ [batches/cycle]\\
                  & batch size & $20$ [MTHM]\\
                  & cycle length & $18$ [months]\\
                  & refuelling time & $2$ [months]\\
                  & requests & recycled MOX (1st preference)\\
                  & requests & enriched UOX (2nd preference)\\
\hline
\textbf{Spent Fuel Fabrication} & requests & depleted uranium\\
\Class{cycamore::Fab} & requests & separated fissile material\\
                      & offers & recycled MOX\\
\hline
\textbf{Separations} & requests & all spent fuel types \\
\Class{cycamore::Separations} & offers & separated fissile material\\
                      & efficiency & 0.99\\
                      & Pu separation capacity & $6.0E4$ [kg/month] \\
\hline
\textbf{Repository} & requests & all waste\\
\Class{cycamore::Sink} & capacity & $\infty$\\
\hline
\textbf{UOX Fabrication} & offers & UOX\\
\Class{cycamore::Source} & capacity & $\infty$\\
\hline
\textbf{DU Source} & offers & depleted U\\
\Class{cycamore::Source} & capacity & $\infty$\\
\hline
\end{tabularx}
\caption{Facility Configurations for Example Simulations}
\label{tab:facconfigs}
\end{table*}
\twocolumn

As described in Table \ref{tab:facconfigs}, the facilities used in these simulations are:
\begin{itemize}

    \item \textbf{Reactor} (\Class{cycamore::Reactor}): This is a reactor
        model that requests any number of input fuel types and transmutes them
        to static compositions when they are burnt and discharged from the
        core. The reactor was configured to model a light water reactor
        with a 3 batch core operating on an 18 month
        cycle with a 2 month refuel time.  The batches in the core each contain 
        20 \gls{MTHM}, where heavy metals are actinide elements like thorium, 
        uranium, and plutonium. The reactor was also configured to
        take in either enriched \gls{UOX} fuel or recycled \gls{MOX} fuel.

    \item \textbf{Spent Fuel Fabrication} (\Class{cycamore::Fab}): This
        facility requests depleted uranium and separated fissile material and
        mixes them together into recycled \gls{MOX} fuel that it offers to
        requesters.  The two streams are mixed in a ratio in order to match
        simple neutronics properties of the target fuel as closely as
        possible.  The method used is based on a variation ``equivalence
        method'' originally developed by Baker and Ross
        \cite{baker_comparison_1963}.  This technique has also been used in the
        \gls{COSI} fuel cycle simulator developed by the French \gls{CEA}.

    \item \textbf{Separations} (\Class{cycamore::Separations}): This facility
        takes in all kinds of spent fuel and separates it into plutonium and
        waste streams with some efficiency (0.99 was used for these
        simulations).  Up to 60,000 kg of Pu can be separated per month.

    \item \textbf{Repository} (\Class{cycamore::Sink}): This is an
        infinite-capacity facility that can take in all types of material
        including separations waste streams and spent reactor fuel.

    \item \textbf{UOX Fabrication} (\Class{cycamore::Source}): This facility
        provides enriched \gls{UOX} fuel as requested.  This facility has infinite
        production capacity.

    \item \textbf{DU Source} (\Class{cycamore::Source}): This facility
        provides depleted uranium as requested. This facility has infinite
        production capacity.

\end{itemize}

To model each of the three fuel cycles, only simple 
adjustments to the input file specification were necessary. Specifically,  
only the commodity types and trade preferences for
each of the facilities needed to be altered from one simulation to another. 

For all cases, the reactor was configured to request recycled \gls{MOX}
fuel with a higher preference than new \gls{UOX} fuel. For the No Recycle case,
the \Class{Reactor} was set to offer all spent fuel as waste.  For the 1-pass
recycle case, the \Class{Reactor} offered spent \gls{UOX} into a spent fuel market,
but spent \gls{MOX} is still offered as waste.  In the Infinite-Pass Recycle case,
the \Class{Reactor} offers all burned fuel into a spent fuel market. The
separations facility requests spent fuel with a higher preference than the
repository resulting in preferential recycle.  All these preferences are easy
to adjust and \Cyclus dynamically handles supply constraints and non-uniform
preference resolution.  It is notable that separations and recycle fuel
fabrication capacity are deployed identically in all simulations.  In the once
through case, the recycling loop never acquires material, and so reactors
always receive fresh \gls{UOX} fuel.  The \gls{DRE} ensures everything operates
smoothly in all cases.

\begin{figure}
\begin{center}
\includegraphics[width=\columnwidth]{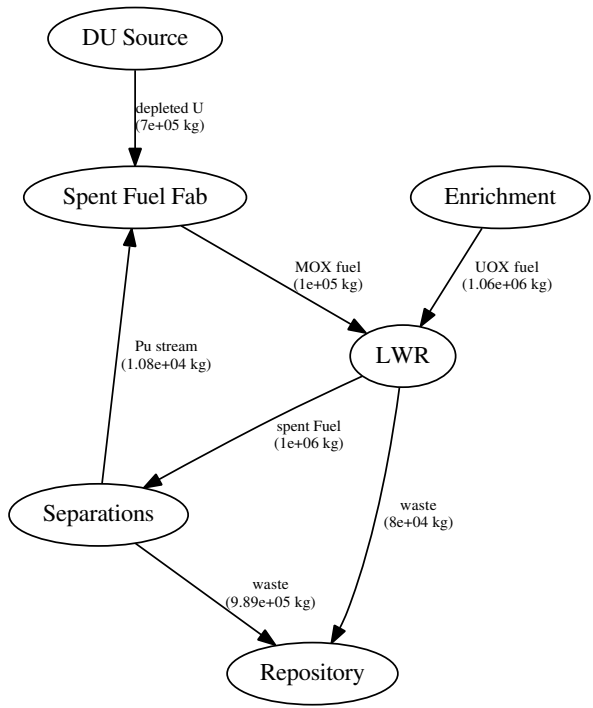}
\end{center}
\caption{1-pass \gls{MOX} recycle scenario material flows.}
\label{fig:flowmodopen}
\end{figure}

\Cyclus' discrete materials make single-pass recycle straightforward to implement.  The
reactors keep track of fuel as discrete batches. A reactor remembers where it
received each batch from.  If a batch was received from the recycled fuel
fabrication facility, it does not offer it to separations and instead offers
it as a waste commodity which is only requested by the repository.  The
material flows for the Single-Pass Recycle case are shown in Figure
\ref{fig:flowmodopen}.

\begin{figure}
\begin{center}
\includegraphics[width=\columnwidth]{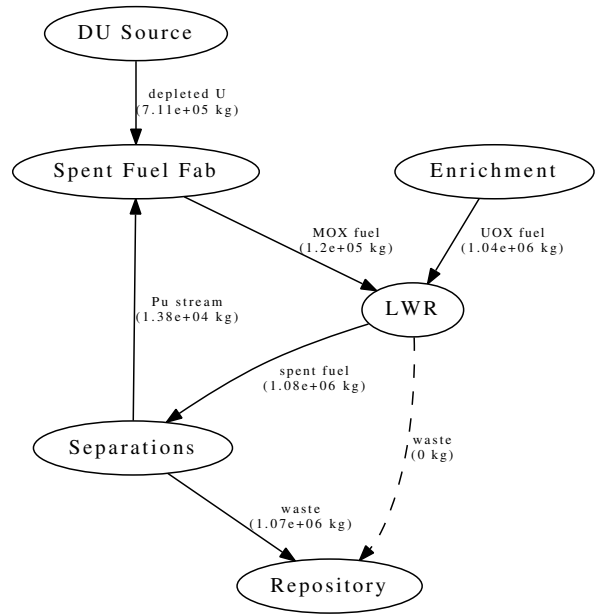}
\end{center}
\caption{Full \gls{MOX} recycle (multi-pass) fuel cycle material flows.}
\label{fig:flowclosed}
\end{figure}

Changing the scenario from a 1-pass fuel cycle to an infinite-pass fuel cycle
requires only a one-word change in the input file regarding the output
commodity for the spent \gls{MOX} fuel of the \Class{Reactor}:

\begin{lstlisting}[language=diff]
  <fuel>
    <incommodity>mox</incommodity>
-   <outcommodity>waste</outcommodity>
+   <outcommodity>spent_fuel</outcommodity>
    <inrecipe>mox_fresh_fuel</inrecipe>
    <outrecipe>mox_spent_fuel</outrecipe>
  </fuel>
\end{lstlisting}

This results in the material flows in Figure \ref{fig:flowclosed}.  A
similarly trivial change was used to switch from the No Recycle to a 1-pass
fuel cycle.  Note that because the reactors always transmute fuel into fixed
compositions, the error in isotopic compositions is larger for the
Infinite-Pass Recycle case.

\begin{figure}[htb]
\begin{center}
\includegraphics[width=\columnwidth]{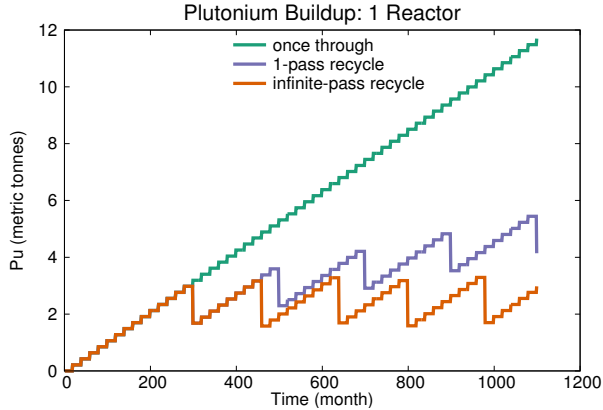}
\end{center}
\caption{System plutonium buildup with one reactor.}
\label{fig:puseries1}
\end{figure}

Figure \ref{fig:puseries1} shows the full-system plutonium buildup for No 
Recycle (once through), Single-Pass Recycle, and Infinite-Pass Recycle variations of
the one-reactor scenario described above. The figure was generated directly
from \Cyclus output data. After several batch cycles (near month 300) in the
1-pass and infinite-pass cases, enough separated fissile material accumulates
in the fuel fabrication facility to generate a full recycled batch.  When this
batch is transmuted, more plutonium is burned than created.  This results in a
drop in the total fuel cycle system plutonium inventory.  This pattern repeats
roughly every 10 cycles (200 months) for the Single-Pass Recycle case and every 9
cycles (180 months) for the Infinite-Pass Recycle case.  Because the 1-pass
recycle scenario does not re-recycle material, it takes the fabrication
facility 2 cycles longer to accumulate a full batch of fissile material.

Because facilities are represented individually and transact discrete
materials as discrete events, realistic non-uniform patterns in facility
behavior that affect total system behavior are observed using \Cyclus.

\begin{figure}[htb]
\begin{center}
\includegraphics[width=\columnwidth]{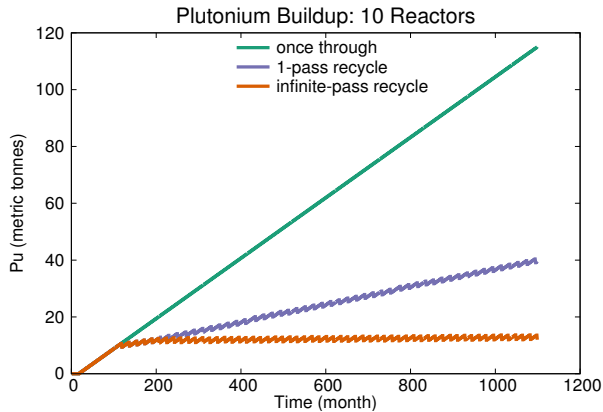}
\end{center}
\caption{System plutonium buildup with staggered refueling for many reactors.}
\label{fig:puseriesn}
\end{figure}

Figure \ref{fig:puseriesn} shows plutonium buildup for the 10-reactor
simulations of the No Recycle, Single-Pass Recycle, and Infinite-Pass Recycle
scenarios.  As the number of reactors (with staggered refueling) increases,
the behavior of the system approaches a more steady average reminiscent of
continuous material flow models.

The fundamental capabilities of demonstrated in these simulations qualify
\Cyclus and its ecosystem to model the breadth of scenario types expected of
nuclear fuel cycle simulator tools. These examples further show the flexibility
provided by the \gls{DRE} logic within the \Cyclus framework and provide
an example of the resolution made possible by discrete-facility and
discrete-material modeling in fuel cycle simulation.

\section{Conclusions}


The \Cyclus nuclear fuel cycle framework presents a more generic and flexible
alternative to existing fuel cycle simulators. Where previous nuclear fuel
cycle simulators have had limited distribution, constrained simulation
capabilities, and restricted customizability, \Cyclus emphasizes an open
strategy for access and development.  This open strategy not only improves
accessibility, but also enables transparency and community oversight. 
Furthermore, the object-oriented \gls{ABM} simulation paradigm ensures more generic
simulation capability. It allows \Cyclus to address common analyses in a more
flexible fashion and enables analyses that are impossible with system dynamics
simulators.

Similarly, the fidelity-agnostic, modular \Cyclus architecture facilitates simulations
at every level of detail. Simulations relying on arbitrarily complex isotopic
compositions are possible in \Cyclus, as are simulations not employing any
physics at all. Physics is introduced through optional system-wide radioactive 
decay of materials and through the use of physics-enabled facility libraries.  
To support calculation of physical processes in nuclear facilities, the 
\Cycamore library provides models employing basic physics for core fuel cycle 
facilities and extension libraries from the community support more detailed 
simulations. Indeed, agents of such varying fidelity can even exist in the same 
simulation. Researchers no longer need to reinvent the underlying simulator 
framework in order to model a simulation focused on the aspects of the fuel 
cycle relevant to their research.

Furthermore, when the capabilities within \Cyclus, \Cycamore, and the rest of
the ecosystem are insufficient, adding custom functionality is simplified by a
modular, plug-in architecture. A clean, modern \gls{API} simplifies
customization and independent archetype development so that researchers can
create models within their domain of expertise without modifying the core
simulation kernel. Throughout the \Cyclus
infrastructure, architecture choices have sought to enable cross-institutional
collaboration and sustainable, community-driven development. The ecosystem
of capabilities, already growing, may someday reflect the full diversity of use
cases in the nuclear fuel cycle simulation domain.

\section{Acknowledgements}
This work has been supported by a number of people. The authors
would like to thank software contributors Zach Welch and Olzhas Rakhimov.
Additionally, the authors are grateful to the many enthusiastic members of the
\Cyclus community and our \gls{NEUP} collaboration partners at the University
of Texas, University of Utah, and University of Idaho.

Direct support for the \Cyclus project has been received from the \gls{NRC}
Faculty Development program, the \gls{NEUP} Research \& Development Program,
the \gls{NNSA} Consortium for Verification Technology, and the \gls{UW}.  In
addition students contributing to \Cyclus have received fellowship support
from the \gls{ANL} Lab Grad program, the \gls{NEUP} Fellowship program, the
\gls{NSF} Graduate Fellowship program, and the \gls{DHS} Nuclear Forensics
Graduate Fellowship program.

\bibliography{fundamentals}

\end{document}